\documentclass[12pt,preprint]{emulateapj}
\usepackage{color}
\usepackage{ulem}

\newcommand{\msun}{\ensuremath{M_\odot}}

\newcommand{\feoh}{{\rm [Fe/H]}}

\newcommand{\omcen}{$\omega$ Cen}

\newcommand{\mmix}{\ensuremath{\delta M_{{\rm mix}}}}
\newcommand{\tmix}{\ensuremath{\tau_{{\rm mix}}}}
\newcommand{\mini}{\ensuremath{M_{{\rm ini}}}}
\newcommand{\yenv}{\ensuremath{Y_{{\rm env}}}}

\submitted{Recieved 2007 August 31; accepted 2007 October 27}

\shorttitle{He-Enhancement in Globular Clusters}
\shortauthors{Suda et al.}

\begin{document}

\title{ Highly He-Rich Matter Dredged Up by Extra Mixing through Stellar Encounters in Globular Clusters}

\author{Takuma Suda$^1$, Takuji Tsujimoto$^2$, Toshikazu Shigeyama$^3$, and Masayuki Y. Fujimoto$^4$}
\altaffiltext{1}{Research Center for the Early Universe, Graduate School of Sciecne, University of Tokyo, 7-3-1 Hongo, Bunkyo-ku, Tokyo 113-0033, Japan; suda@astro1.sci.hokudai.ac.jp; Current address: Department of Physics, Hokkaido University, N10 W8, Kita-ku, Sapporo, 060-0810, Japan}
\altaffiltext{2}{National Astronomical Observatory, Mitaka-shi, Tokyo 181-8588, Japan; taku.tsujimoto@nao.ac.jp}
\altaffiltext{3}{Research Center for the Early Universe, Graduate School of Sciecne, University of Tokyo, 7-3-1 Hongo, Bunkyo-ku, Tokyo 113-0033, Japan; shigeyama@resceu.s.u-tokyo.ac.jp}
\altaffiltext{4}{Department of Physics, Hokkaido University, N10 W8, Kita-ku, Sapporo, 060-0810, Japan; fujimoto@astro1.sci.hokudai.ac.jp}

\begin{abstract}
The unveiled main-sequence splitting in $\omega$ Centauri as well as NGC 2808 suggests that matter highly-enriched in He (in terms of its mass fraction $Y\sim0.4$) was produced and made the color of some main-sequence stars bluer in these globular clusters (GCs). The potential production site for the He-rich matter is generally considered to be massive AGB stars that experience the second dredge-up. However, it is found that massive AGB stars provide the matter with
$Y\sim 0.35$ at most, while the observed blue-shift requires the presence
of $Y\sim 0.4$ matter. Here, we show that extra mixing, which operates in the red giant phase of stars less massive than $\sim2\,M_{\odot}$, could be a mechanism that enhances He content in their envelopes up to $Y\sim 0.4$. The extra mixing is supposed to be induced by red giant encounters with other stars in a collisional system like GCs.  The $Y\sim 0.4$ matter released in the AGB phase has alternative fates to (i) escape from a GC or (ii) be captured by kinematically cool stars through encounters. The AGB ejecta in $\omega$ Cen, which follows the latter case, can supply sufficient He to cause the observed blue-shift. Simultaneously, this scheme generates the extreme horizontal branch, as observed in $\omega$ Cen in response to the higher mass loss rates, which is also caused by stellar encounters.
\end{abstract}


\section{Introduction}

The discovery of a split in the main sequence (MS) of \omcen\  \citep{Bedin04} has opened a new window in the study of Galactic globular clusters (GCs). The observed fact that stars of the blue MS (bMS) in \omcen\ exhibit slightly stronger absorption of iron lines on average, in comparison with the red MS (rMS) \citep{Piotto05} which strongly implies that the origin of bMS is attributable to the enhancement of He inside bMS stars. Subsequently, the recent HST observation has provided us with the second sample of MS splitting, NGC 2808 \citep{Piotto07}.  This GC essentially exhibits no dispersion of [Fe/H] and thus the argument of He enhancement in bMS stars has been reinforced. A comparison between the observed color-magnitude diagrams (CMDs) and synthetic population models suggests that bMS He abundance is enhanced to $Y\sim 0.4$ for both of two GCs \citep{Norris04, Lee05, DAntona05, Piotto07}. \citet{Tsujimoto07} have argued that if the surface of a star on the MS is polluted with  the $Y=0.4$ matter with the mass of $\sim 0.1 \msun$, the star can move to a position on the observed bMS in the CMD. This relaxes a severe demand on the amount of He possibly supplied from asymptotic giant branch (AGB) stars, in contrast to the idea that the bMS stars are born from pure AGB ejecta consisting of $Y\sim 0.4$.
Similarly, \citet{Newsham07} have insisted that surface pollution may explain the high He content of bMS stars.

It should be stressed that $Y\sim 0.4$ matter is necessary to realize the bMS stars in these GCs. Since massive AGB stars can enhance the He abundance in their envelopes through the second dredge-up in the early AGB phase, they have been considered a possible production site for the $Y\sim 0.4$ matter \citep{DAntona05}. However, previous studies indicated that the resultant abundance of He in the envelope of any AGB star does not exceed $Y \sim 0.35$ \citep[see e.g.,][]{vandenHoek97}. Therefore, such a He yield from massive AGB stars may not be sufficient to split the MS as observed. In the subsequent section, we will discuss this issue and conclude that it is highly implausible for a massive AGB star to enhance He abundance up to $Y\sim 0.4$ in its envelope. 

Then, we should search for an alternative mechanism that enhances the He abundance more efficiently than the second dredge-up. We believe that this process, which enhances He abundance, should be accompanied by some other elemental signatures. This is reminiscent of the abundance anomaly of red giants in GCs. The abundance anomalies observed for red giants, such as the abundance variations of CNO elements and the O-Na anticorrelation, are common attributes associated with GCs including \omcen\  \citep[see][]{Gratton04}. One of the proposed mechanisms to explain these anomalies is a deep mixing, so-called extra mixing,  on the red giant stage \citep{Sweigart79, Langer93, Charbonnel98, Fujimoto99,Aikawa01,Chaname05,Suda06}.
The extra mixing is assumed to be caused by some kind of rotation-induced mixing, driven by shear instability around the border of the He core \citep[see][]{Fujimoto88,Zahn92}. Since such a deep mixing naturally draws He from the core, the extra mixing is a promising mechanism of producing He-rich matter.
As a mechanism of mixing, most of the previous works assumed a continuous mixing caused by the meridional circulation, while Fujimoto and his coworkers insisted that a flash of H brought into a degenerate core by the rotation-induced mixing drives the intermittent mixing.

In this study, we propose that extra mixing, which occurs in the red giant phase of a star, enhances He abundance in the envelope and ejects He-rich matter when the star evolves to a white dwarf through mass loss. In this scenario, the driving force of the extra mixing is generated through interactions between the red giant and dwarfs in the central region of \omcen. Our detailed calculations reveal that stars with the masses less than $\sim$2 \msun\ can enhance He abundance up to $Y\sim 0.4$, provided that the extra mixing operates in the red giant phase. 

Based on our scenario, we discuss other conspicuous characteristics of \omcen, the extremely blue horizontal branch (HB), which is observed in some other Galactic GCs.
Although factors that determine HB morphology is still an open question and is known as the second parameter problem, the most effective parameter that affects the location of a HB star in the CMD is the mass $M_{\rm env}$ of the envelope.
A major factor to determine $M_{\rm env}$ is the mass loss during the red giant phase.
We predict that encounters induce a high mass loss rate through a gain of angular momentum and promote the presence of an extremely blue HB population. In the end, two mysteries in \omcen\  of MS splitting and HB morphology are discussed in a unified framework of encounters of dwarfs with AGB ejecta and those of red giants with dwarfs. 

\section{Massive AGB stars as a source of He-rich matter}

In this section, we estimate He abundance in the envelope of a massive AGB star after the second dredge-up and show that it is enhanced as high as $Y \sim 0.35$ but cannot significantly exceed this value. The He enhancement $\Delta \yenv$ in the envelope due to the second dredge-up is determined by (1) the envelope mass, (2) the He core mass $M_{\rm core}$ just before the second dredge-up, and (3) the maximum depth of the convective envelope during the second dredge-up whose mass coordinate is given by $M_{\rm mix,\,base}$. For a massive AGB star, the envelope mass is not reduced significantly by the mass loss until the beginning of thermal pulses. The surface He abundance becomes smaller for a more massive envelope since the matter dredged-up from the He core is diluted in the whole convective envelope. Assuming a simple stellar structure consisting of the He core and the H-rich envelope, the He enhancement $\Delta \yenv$ in the envelope is approximated by the following formula;
\begin{equation}
\Delta \yenv = \frac{(M_{\rm core} - M_{\rm mix,\,base})(1 - \yenv)}{M_{\rm init} - M_{\rm mix,\,base}}  \ \ ,
\label{eq:yenv}
\end{equation}
where $M_{\rm init}$ denotes the initial mass of a star. The core mass $M_{\rm core}$ is calculated with the stellar evolution program for a set of given initial values of mass. The initial value of \yenv\ is assumed to be 0.24 in the rest of this section. Using this formula, we can estimate the amount of the dredged-up matter during the AGB mass loss phase to attain a given enhancement of He. For example, $M_{\rm mix,\,base} \approx 0.32$ \msun\ is required to attain $\Delta \yenv \simeq 0.15$ by the second dredge-up for a star with $M_{\rm init}=7$ \msun\ ($M_{\rm core}=1.68$ \msun) if the envelope does not significantly lose its mass. Of course, it is impossible to dredge-up He from such a deep interior since the C-O core of this particular star becomes as massive as $\sim 0.9$ \msun\ before the second dredge-up. Therefore, this star cannot enhance the He content in the envelope to $\yenv=0.4$. 

After the second dredge-up, the mass of He in the He-burning shell is reduced from inside due to the growth of the C-O core. Thinning of the He-burning shell inhibits  further enhancement of the He content in the envelope by the subsequent third dredge-up. For example, in a 7 \msun\ star, He shell flashes convert some He into C and dredge up the He layer with the mass of 0.1 \msun\  ($Y\simeq0.8$) to the envelope. The He enhancement for a 7 \msun\ star (6 $\msun$ envelope)
is estimated to be only $\Delta Y_{\rm env}\sim0.013$. As a result of the above formula as well as  previous studies on stellar evolution
,He content in the envelope of a massive AGB star with $\mini \gtrsim 3$ \msun\ can attain $\yenv\sim 0.36$ at most and never exceeds 0.4.
Indeed, no established models have been proposed to attain $\yenv\sim 0.4$ in the envelope of a massive AGB star.
As another problem, it has been discussed \citep{Lattanzio04} that the yield of CNO and light $p$-capture elements in massive AGB stars cannot explain the abundance anomalies in GCs.

\section{Stellar models with extra mixing}

Models in this work are computed with a stellar evolution program based on \citet{Iben92},  which includes updated physics and improved opacity tables \citep[][Suda \& Fujimoto 2007, in preparation]{Suda04, Suda07a}. To construct isochrones for MS, we evolve stars through 13 Gyr starting from the zero-age MS for various initial masses at an interval of 0.01 \msun. Each evolutionary point is interpolated from the lower MS to subgiant. On the other hand, the isochrones for the red giant branch (RGB) are substituted by the evolutionary tracks of a star with the appropriate mass, since the isochrones on the RGB coincide with the evolutionary tracks for stars with the same initial masses and metallicities. In this RGB phase, we assume that the mixing event occurs through encounters in a GC and the matter in the core is intermittently dredged up to the envelope and mixed. 

The total amount of He in the envelope is determined by the mixing rate, which is the amount \mmix\ of mixed He per event divided by the interval $\tmix$ between the mixing events.
The duration of the RGB phase should be much longer than the time for a star to cross the cluster core $D/v\sim 1\,{\rm pc}/10 \,{\rm km~s}^{-1}\sim10^5$ yr, where $D$ denotes the size of the cluster core and $v$ the velocity of a star.
While a red giant passes the cluster core, the star often encounters other stars and its envelope acquires some amount of the orbital angular momentum. The resultant excess rotation in the envelope causes mixing through shear instabilities at the boundary between the core and envelope.
There is a critical value for the mixing rate \mmix\ defined by the dredged-up mass per mixing event;
$\mmix / \tmix = \langle L_{\rm H} \rangle / ( E_{\rm H} X_{\rm env} )$,
where  $E_{\rm H}$ denotes the energy per unit mass released by H-burning, $\langle L_{\rm H} \rangle$ the average luminosity of the star, and $X_{\rm env}$ is the H abundance in the envelope. The righthand member increases as helium mixing proceeds.
In our models, the values of \mmix\ are chosen to be near the obtained critical mixing rate.
If \mmix\ is too large relative to the critical rate, the He core cannot grow. 
As He mixing proceeds, however, it grows since the critical rate is a decreasing function of $X_{\rm env}$.
The border of the occurrence of He mixing is assumed to be $M_{\rm ini} = 2.1 \msun$. A star less massive than this has a degenerate He core. Once such a star on the RGB encounters another star, the angular momentum in the envelope increases to induce shear between the core and envelope. Then a fraction of H in the envelope is repeatedly mixed into the core and undergoes the off-center flash. This off-center H flash is believed to induce extra mixing \citep{Fujimoto99}.  In more massive stars, the mixed H would not undergo a flash because electrons in the core are not degenerate.

\section{Results}

\subsection{$Y$ as a function of stellar mass}
The evolutions of He core mass and content $\yenv$ in the envelope during the RGB for a star with $M_{\rm ini}=1.5$\msun\ are shown as an example in Figure 1. 
Basically, a larger \mmix\ leads to greater He enhancement. For the case with \mmix$=0.0015$ \msun,
the core that was once reducing its mass begins to grow after the first several mixing events due to the increasing $\yenv$.
Then, it grows rapidly as the H-burning rate increases with the core mass. The resultant $\yenv$ at the tip of RGB becomes greater than 0.4. This, together with other low-mass stellar models, reveal that stars with $M_{\rm ini}\lesssim 2$\msun\ can achieve $\yenv> 0.4$  at the tip of the RGB as a result of the extra mixing. These He-rich matters are ejected at the final stage of AGB. Figure~2 shows the He abundance $\yenv$ finally attained in the envelope as a function of the initial mass $M_{\rm ini}$ of stars in the range of $0.8\msun \leq M_{\rm ini}\leq 8\msun$, together with the average yield from massive AGB stars with $M_{\rm ini}\geq 3\msun$ by other studies. 
Note that $Y\sim 0.4$ matters are produced from low-mass stars only in GCs where red giants encounter dwarfs.
Additionally, it should be stressed that the He-rich matter is expected to show C depletion and N enhancement as a result of partial CN cycles. For the parameter range surveyed by this work, we obtain [C/Fe] $= -0.2 - -0.5$ and [N/Fe] $= 0.8 - 0.9$. These results are marginally compatible with the observed values [C/Fe] $\sim -0.1 - 0$ and [N/Fe] $\gtrsim 1$ for bMS stars (Piotto et al. 2005) and subgiants with $<$[Fe/H]$>$=$-1.37$ \citep{Villanova07}.

\begin{figure}
\begin{center}
\plotone{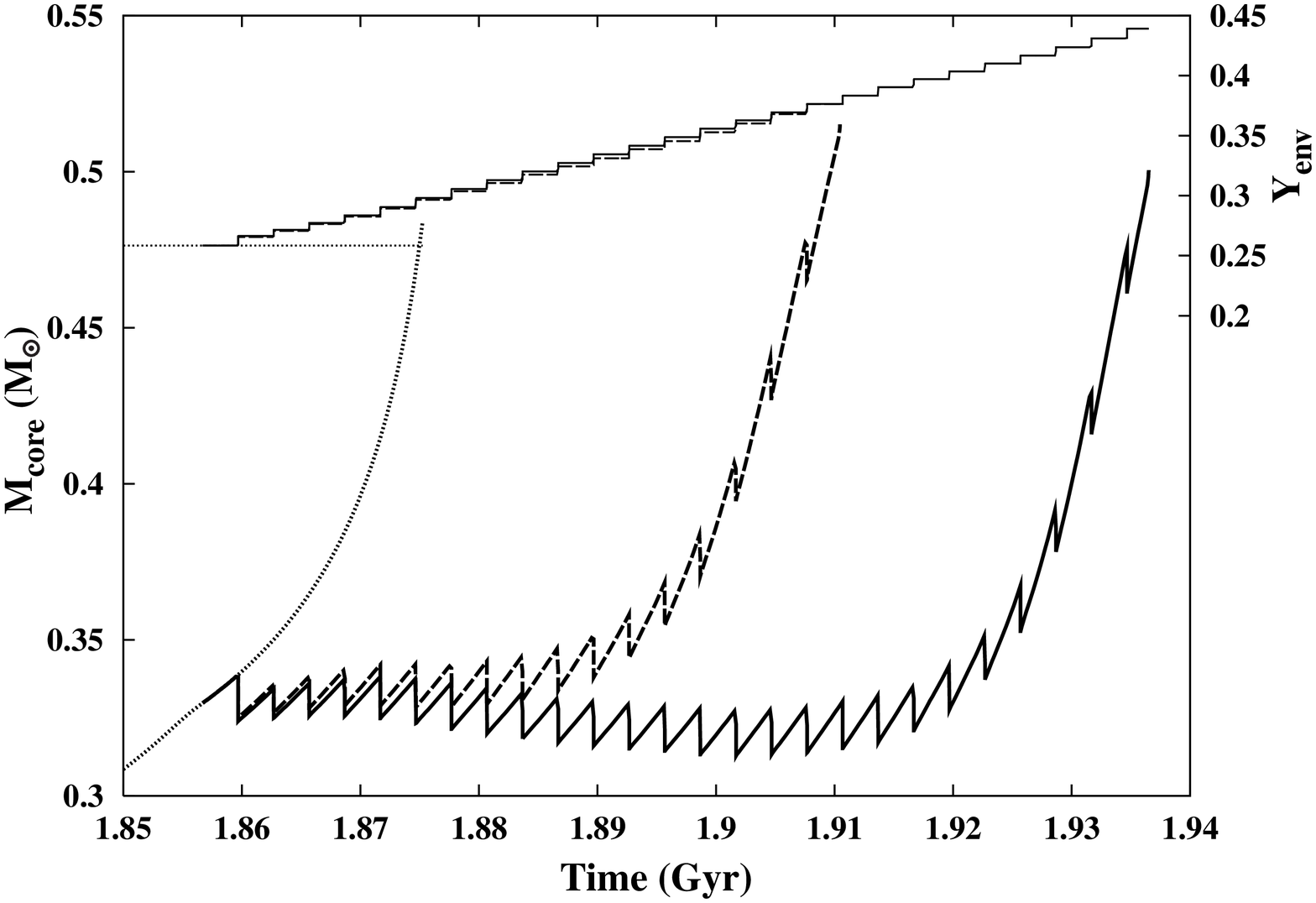}
\vspace{0.5cm}
\caption{
Time evolution of the He core mass $M_{\rm core}$ and the He abundance $\yenv$ in the envelope during the RGB phase for a 1.5 $\msun$ star without  (dotted line) and with extra mixing. In the extra mixing models, $\mmix$ = 0.0014 $\msun$ (dashed line) and 0.0015 $\msun$ (solid line) are assigned. The value of $\tmix$ is set to be $3\times 10^{6}$ yr. The calculations are performed until the tip of the RGB.
}
\end{center}
\label{fig:mcore}
\end{figure}

\begin{figure}
\begin{center}
\plotone{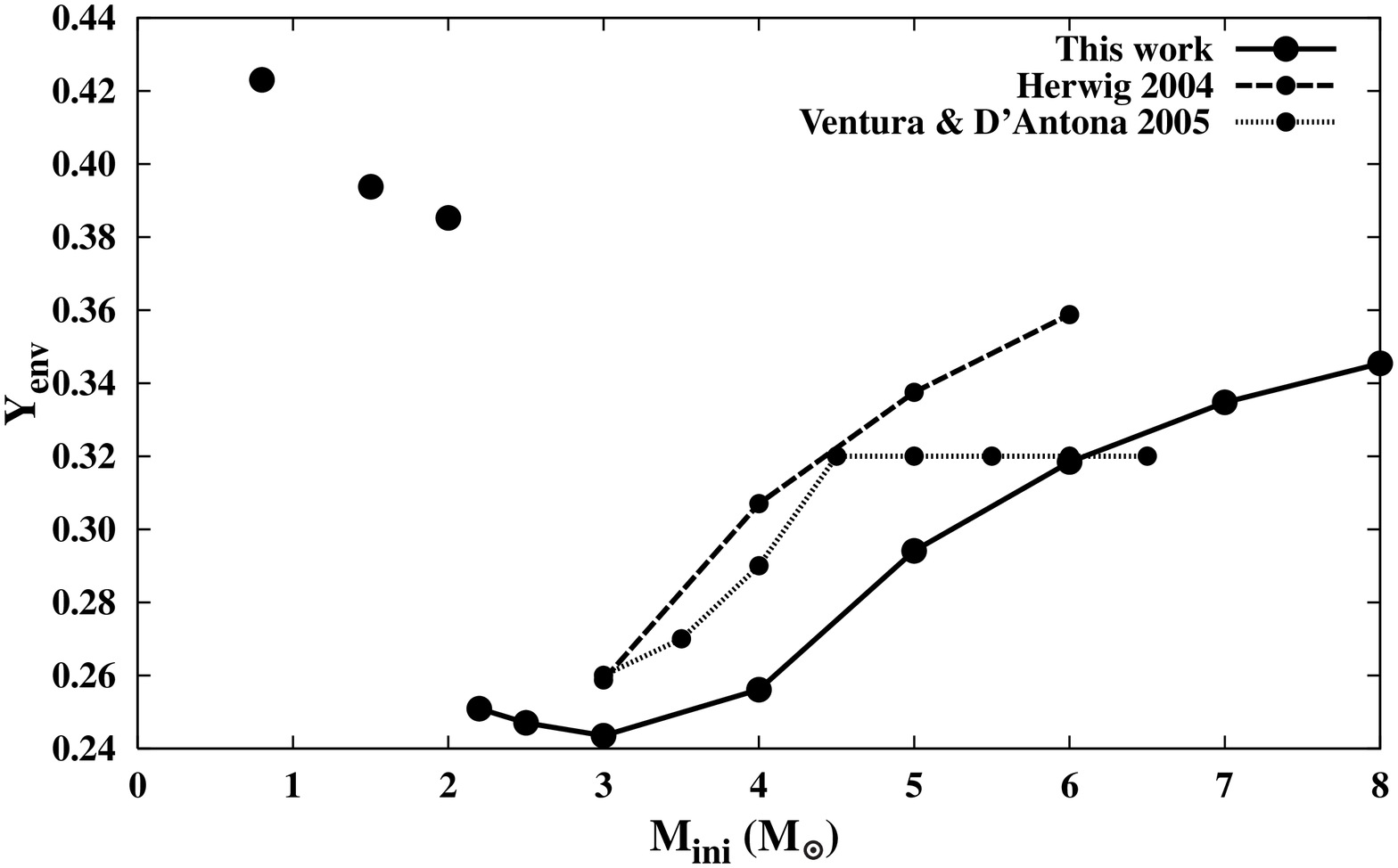}
\caption{He abundance $\yenv$ finally attained in the envelope as a function of the initial stellar mass, $\mini$. For $\mini \leq 2.1 \msun$, extra mixing is assumed to occur during the RGB. The interval of mixing is assumed to be $\tmix=3\times10^{5}$ yr. The values of the mixed mass per event are set to be \mmix $=0.0015,\,0.0025,\,0.0035$ \msun\ , for models with $M_{\rm init}=0.8,\,1.5,\,2.0$ \msun, respectively. Our results for $M_{\rm ini}\geq 3\msun$ are compared with other studies by \citet{Herwig04} and \citet{Ventura05}.}
\end{center}
\label{fig:Y}
\end{figure}

On the other hand, stars with $M_{\rm ini} > 2.1\msun$ enhance the He abundance in their envelopes through the process to deepen a convective envelope by the second dredge-up. Furthermore, stars with a mass of $\sim$3.0 \msun\ stay on the RGB only for $\sim 10^6$ yr. Therefore, the chance of an encounter is so limited that such stars are unlikely to undergo extra mixing, even if the extra mixing did not need a degenerate He core. The $\yenv$ of these massive AGB stars
increases with increasing stellar masses and a $M_{\rm ini}=8$\msun\ star gives $\yenv\sim 0.35$ as the maximum yield. As presented by \citet{Herwig04}, a model adopting overshooting leads to efficient enhancement of $\yenv$ after the second dredge-up. At the same time, however, it accelerates the growth of the AGB core. As a result, the lower mass limit of a star that undergoes central C-burning decreases to $M_{\rm ini} \sim7\msun$ from $\sim8\, \msun$. Therefore, even with overshooting, it is unlikely to realize \yenv\ $\sim$ 0.4 in massive AGB stars.

\subsection{Surface Pollution and Main-Sequence Splitting}

As discussed in \citet{Tsujimoto07}, a 0.1\msun\ accretion of $Y=0.4$ matter by each star is sufficient to reproduce a bifurcation of the MS as observed in \omcen. Thus, the total He mass supposed to be accreted by the bMS stars is estimated to be $\sim 0.012 M_\omega$, where $M_\omega$ denotes the stellar mass of \omcen, if the mass fraction of the bMS stars is assumed to be 0.25. On the other hand, the amount of He ejected from low mass AGB stars of all generations of stars is calculated to be $0.02 M_\omega$, which is greater than the former estimate and thus satisfies a necessary condition on the supply and demand of He. Figure~3 shows the CMD of the 13 Gyr isochrones for the MS and RGB calculated with two models, one with the accretion of He-rich matter  (\feoh = -1.3, $Y$=0.24) and the other without the accretion (\feoh = -1.6, $Y$=0.24).
Here, we assume that each star accretes a $Y=0.4$ matter at a rate of 0.01\msun\ Gyr$^{-1}$ during 1-11 Gyr. For reference, stars composed of a $Y=0.4$ gas are also plotted. It is shown that a $Y=0.4$ matter either on the surface (solid line) or an entire star (dotted line) shifts the color of main-sequence to bluer, as compared to a normal $Y=0.24$ star (dashed line). 

\begin{figure}
\begin{center}
\plotone{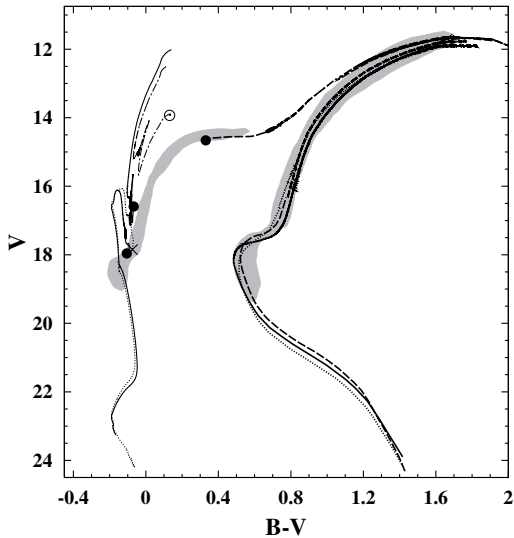}
\caption{Theoretical CMDs predicted by our three models, i.e., the surface pollution with extra mixing model (solid line), $Y=0.24$ model (dashed line) and $Y=0.4$ model (dotted line) together with the observed CMD for \omcen, which is indicated by the shaded area \citep{Rey04}. The theoretical CMDs are presented by isochrones at the age of 13 Gyr for MS and RGB, and evolutionary tracks for HB. The positions of ZAHB are shown by various symbols. The extra mixing models with an efficient mass loss are denoted by lower two filled circles, attached with evolutionary tracks (solid lines), whereas an open circle is for the model with the normal mass loss estimated from $\eta = 1/3$ in Reimers' formula, together with an evolutionary track (dot-dashed line). A cross with a dotted track is for the $Y=0.4$ model, and the other filled circle at V$\sim 15$ expresses the $Y=0.24$ model. The three ZAHBs in the extra mixing model have total masses of 0.61, 0.55, and 0.52 $\msun$, respectively, from top to bottom. Here, we adopt the distance modulus $(m - M)_{\textrm v} = 13.5$ and the reddening $E(B-V) = 0.14$.
}
\end{center}
\label{fig:cmd}
\end{figure}

\subsection{Horizontal Branch Morphology}

For later evolution, evolutionary tracks are drawn for HB in the CMD from the tip of the RGB. In the surface pollution model, extra mixing is assumed to operate during the RGB and the calculation is performed for a $0.8 \msun$ star, whereas a $0.64\, \msun$ star is considered for the $Y=0.4$ model because a He-rich star has a short lifetime. Because of its small mass, the envelope mass of a $Y=0.4$ HB star is reduced sufficiently to populate an extremely blue HB (cross), as dictated by the assumption of the standard mass loss formula ($\eta = 1/3$)\citep{Reimers77}. On the other hand, the zero-age HB (ZAHB) of a star, in which the extra mixing operated during the RGB, is located less bluer and brighter in the horizontal part of HB (open circle), as long as the standard mass loss rate is assumed. However, our scenario predicts that excess angular momentum in the envelope acquired by stellar encounters will enhance the mass loss rate. Thus, a star that experienced mixing is expected to have a reduced envelope mass and its predicted location of ZAHB becomes close to the observed extreme HB, as shown by the lower two filled circles in Figure~3. Such an efficient mass loss is supported by the existence of dim stars around $V\sim 19$ mag in the observed CMD \citep{Rey04}. Our model predicts that the positions of these stars in the CMD are in the pathway of stars that extinguish the H shell-burning during the HB evolution due to their decreasing envelopes, and finally move to the white dwarf branch. Figure 4 summarizes the dependence of the ($B-V$) color of ZAHB on the envelope mass for each model.

\begin{figure}
\begin{center}
\plotone{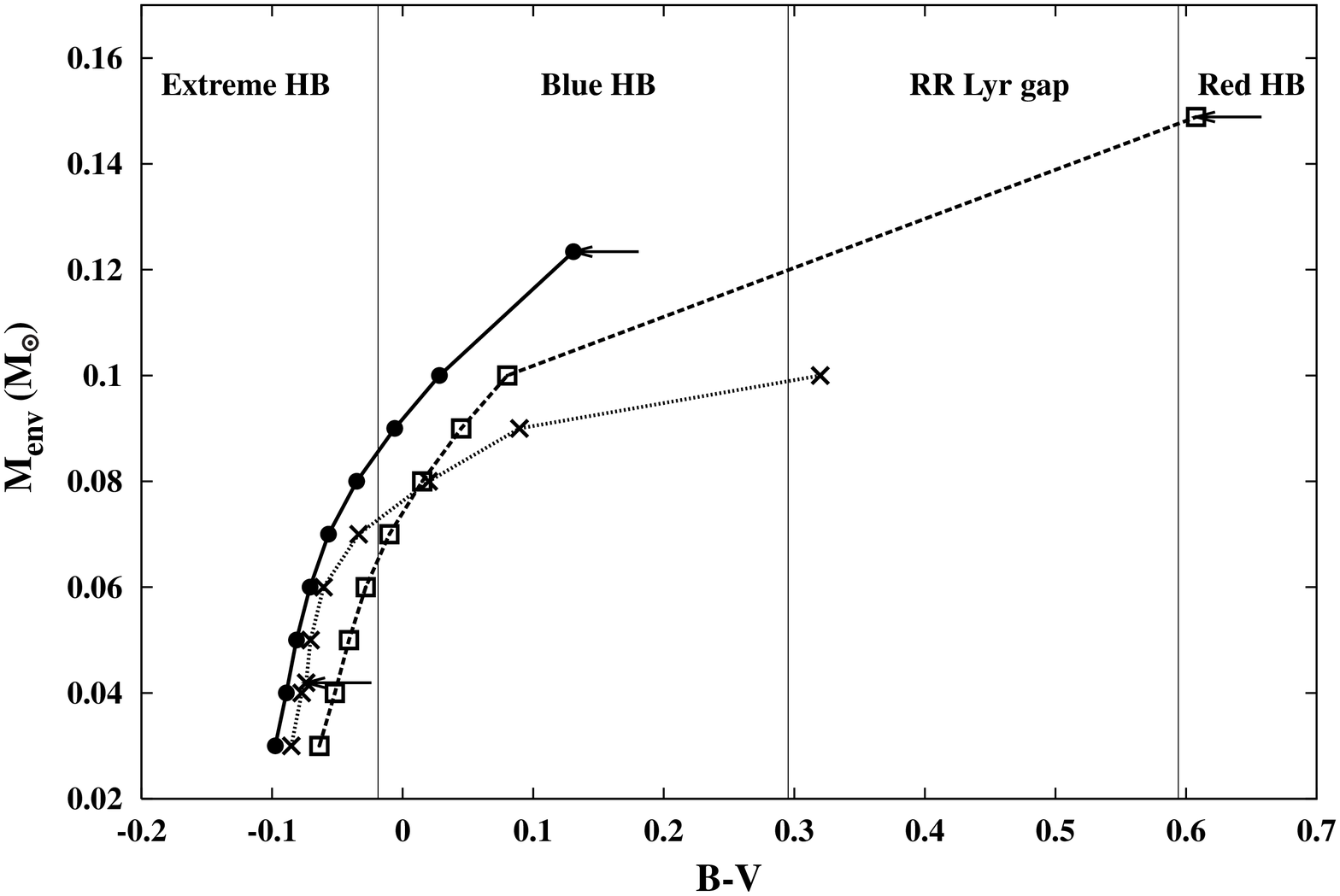}
\caption{The relationship between the (B-V) color of ZAHB and the envelope mass $M_{\rm env}$ for a fixed He core mass for each of the three models (extra mixing model with \yenv\ $=0.42$: solid line, $Y=0.24$: dashed line, $Y=0.4$: dotted line). The He core mass for each model is set to be 0.4923, 0.4861, and 0.4632 \msun, respectively. In each model, the results with $\eta = 1/3$ in Reimers' mass loss formula are included and indicated by arrows. }
\end{center}
\label{fig:mloss}
\end{figure}

\section{Conclusions}
We first propose that He-rich matter is synthesized in red giants with the masses of 0.8 - 2 $\msun$, which experience encounters with other stars in GCs, leading to the extra mixing during the RGB phase in their envelopes. On the other hand, massive AGB stars are unable to produce the He abundance as much as $Y \sim 0.4$ because (1) their large envelopes work as a buffer against He enrichment and (2) the amount of dredged-up matter is limited by thinning the He-burning shell, due to the growth of C-O core. Therefore, a $Y\sim 0.4$ matter is a product unique to GCs. In some GCs, this matter is retained in their gravitational potential for a prolonged time and is accreted by kinematically cool stars, although this may not be the case for most GCs. In addition, stellar encounters in GCs induce the extra mixing and increase the mass loss rate during the RGB phase. As a result, such stars are predicted to evolve to significantly blue HB stars. In conclusion, MS splitting is inclined to be accompanied by the existence of an extremely blue HB, as observed in \omcen\ and NGC 2808.
It should be, however, noted that our models might make a variation in the final He abundance as a result of extra mixing induced by various amounts of angular momentum transfered through encounters. If the variation is too large, then it will erase the MS splitting and lead to a broad MS. Further investigations of the extra mixing model in terms of the MS splitting are surely awaited.

\acknowledgements
This work is in part supported by a Grant-in-Aid for Science Research from the Japansese Society for the Promotion of Science (19740098).



\begin{thebibliography}{}
\bibitem[Aikawa et al.(2001)]{Aikawa01} Aikawa, M., Fujimoto, M. Y., \& Kat$\bar{o}$, K. 2001, \apj, 560, 937
\bibitem[Bedin et al.(2004)]{Bedin04} Bedin, L. R., Piotto, G., Anderson, J., Cassisi, S., King, I. R., Momany, Y., \& Carraro, G. 2004, \apjl, 605, L125
\bibitem[Charbonnel at al.(1998)]{Charbonnel98} Charbonnel, C., Brown, J. A., \& Wallerstein, G. 1998, \aap, 332, 204
\bibitem[Chaname at al.(2005)]{Chaname05} Chaname, J., Pinsonneault, M., \& Terndrup, D. M. 2005, \apj, 631, 540
\bibitem[D'Antona et al.(2005)]{DAntona05} D'Antona, F., Bellazzini, M.,  Caloi, V., Fusi Pecci, F., Galleti, S., \& Rood, R. T. 2005, \apj, 631, 868
\bibitem[Fujimoto(1988)]{Fujimoto88} Fujimoto, M. Y. 1988, \aap, 198, 163
\bibitem[Fujimoto et al.(1999)]{Fujimoto99} Fujimoto, M. Y., Aikawa, M., \& Kat$\bar{o}$, K. 1999, \apj, 519, 733
\bibitem[Gratton et al.(2004)]{Gratton04} Gratton R., Sneden, C., \& Carretta, E. 2004, \araa, 42, 385
\bibitem[Herwig (2004)]{Herwig04} Herwig F. 2004, \apjs, 155, 651
\bibitem[Iben et al.(1992)]{Iben92} Iben, I.~J., Fujimoto, 
M.~Y., \& MacDonald, J.\ 1992, \apj, 388, 521 
\bibitem[Langer et al.(1993)]{Langer93} Langer, N., Hoffman, R., \& Sneden, C., 1993, \pasp, 105, 301
\bibitem[Lattanzio et al.(2004)]{Lattanzio04} Lattanzio, J., Karakas, A., Campbell, S., Elliott, L., \& Chieffi, A. 2004, MmSAI, 75, 322
\bibitem[Lee et al.(2005)]{Lee05} Lee, Y-W., et al.\ 2005, \apjl, 621, L57
\bibitem[Newsham \& Terndrup (2007)]{Newsham07} Newsham, G., \& Terndrup, D. M. 2007, \apj, 664, 332
\bibitem[Norris(2004)]{Norris04} Norris, J. E. 2004, \apjl, 612, L25
\bibitem[Piotto et al.(2005)]{Piotto05}Piotto, G., et al.\ 2005, \apj, 621, 777
\bibitem[Piotto et al.(2007)]{Piotto07}Piotto, G., et al.\ 2007, \apj, 661, L53
\bibitem[Reimers(1977)]{Reimers77} Reimers, D. 1977, \aap, 61, 217
\bibitem[Rey et al.(2004)]{Rey04} Rey, S.-C., Lee, Y.-W., Ree, C. H., Joo, J.-M., \& Sohn, Y.-J. 2004, \aj, 127, 958
\bibitem[Suda et al.(2004)]{Suda04} Suda, T., Aikawa, M., Machida, M. N., Fujimoto, M. Y., \& Iben, I. Jr. 2004, \apj, 611, 476
\bibitem[Suda \& Fujimoto(2006)]{Suda06}Suda, T. \& Fujimoto, M. Y. 2006, \apj, 643, 897
\bibitem[Suda et al.(2007)]{Suda07a}Suda, T., Fujimoto, M. Y., \& Itoh, N. 2007, \apj, 667, 1206
\bibitem[Sweigart \& Mengel(1979)]{Sweigart79} Sweigart, A. V., \& Mengel, J. G. 1979, \apj, 229, 624
\bibitem[Tsujimoto et al.(2007)]{Tsujimoto07}Tsujimoto, T., Shigeyama, T., \& Suda, T. 2007, \apjl, 654, L139
\bibitem[van den Hoek \& Groenewegen(1997)]{vandenHoek97}van den Hoek, L. B., \& Groenewegen, M. A. T. 1997, \aaps, 123, 305
\bibitem[Ventura \& D'Antona (2005)]{Ventura05} Ventura, P., \& D'Antona, F. 2005, \aap, 439, 1075
\bibitem[Villanova et al.(2007)]{Villanova07} Villanova, S., et al. 2007, \apj, 663, 296
\bibitem[Zahn(1992)]{Zahn92} Zahn, J.-P. 1992, \aap, 265, 115
\end{thebibliography}
\end{document}